# Hybrid Heterogeneous Routing Scheme for Improved Network Performance in WSNs for Animal Tracking

Trupti. M. Behera, Sushanta. K. Mohapatra, Umesh. C. Samal, Mohammad. S. Khan

School of Electronics Engineering, KIIT University, Bhubaneswar, Odisha, India.
(e-mail: truptifet@kiit.ac.in ; skmctc74@gmail.com ; umesh.samalfet@kiit.ac.in)
Department of Computer & Information Sciences, East Tennessee State University, Johnson City, USA.
(e-mail: adhoc.khan@gmail.com ).

## Abstract

Wireless Sensor Networks (WSNs) experiences several technical challenges such as limited energy, short transmission range, limited storage capacities, and limited computational capabilities. Moreover, the sensor nodes are deployed randomly and massively over an inaccessible or hostile region. Hence WSNs are vulnerable to adversaries and are usually operated in a dynamic and unreliable environment. Animal tracking using wireless sensors is one such application of WSN where power management plays a vital role. In this paper, an energy-efficient hybrid routing method is proposed that divides the whole network into smaller regions based on sensor location and chooses the routing scheme accordingly. The sensor network consists of a base station (BS) located at a distant place outside the network, and a relay node is placed inside the network for direct communications from nodes nearer to it. The nodes are further divided into two categories based on the supplied energy; such that the ones located far away from BS and relay have higher energy than the nodes nearer to them. The network performance of the proposed method is compared with protocols like LEACH, SEP, and SNRP, considering parameters like stability period, throughput and energy consumption. Simulation result shows that the proposed method outperforms other methods with better network performance.

## 1. Introduction

In WSN, sensor nodes or motes are deployed either deterministically or randomly in the region of interest; to monitor different environmental or physical condition [1]. A sensor node is a battery-driven wireless device that communicates the sensed information to the end user through a base station (BS) or sinks. The major components of sensor nodes are shown in Figure 1:
  a. *Sensing Unit:* It consists of one or a multiple numbers of sensors and an Analog to Digital Converters (ADC).
  b. *Processing unit:* It receives the signal from the sensing unit and manages the collaboration of sensor nodes to carry sensing tasks. It consists of a microprocessor or microcontroller with memory. Intelligent control of the sensor nodes is also done by this unit.
  c. *Communication or Transceiver Unit:* It is responsible for data transmission and reception over a radio frequency (RF) channel. It also interconnects the nodes to the network.
  d. *Power unit:* It generally consists of a battery. The power required to operate all the components of the system is supplied by this unit.

There may be some other units that are application dependent. All the units are integrated to form a tiny module considering its low production cost and low power consumption.

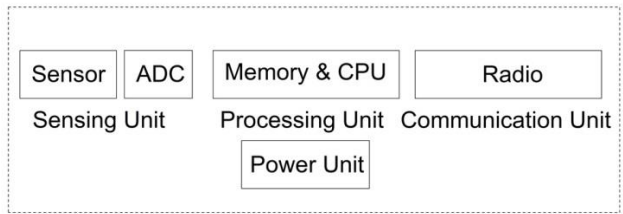
Figure 1: Structure of a Sensor Node

One of the significant design issues in WSN is the efficient management of power. It can be done by proper use of energy at each level of communication starting from data collection to data transmission and reception. From the literature, it has been found that the clustering based routing protocols have the ability to use power efficiently. Clustering algorithms have been the best choice to reduce power consumption in the network level [2]. It is a method to group the nodes into clusters, where each cluster is headed by a cluster head (CH) [3]. The sensing nodes will need enormous power to transfer data directly to BS instead they send it to the nearby CH who also takes the responsibility to aggregate and group the data before sending to BS. Low energy adaptive clustering hierarchy (LEACH) [4][5] is the primitive clustering protocol that distributes energy dissipation evenly amid all nodes within the network. The CH allocates TDMA [6] slots for the member nodes to transmit its sensed data only during its defined time. The node turns off its radio during idle condition. This TDMA scheduling saves energy and extends the network lifetime. Further, the CH fuses the data to filter out any redundant information and then transmits to the BS. Sensor networks placed at distant geographical locations as in the case of animal tracking should be able to tolerate critical conditions to survive long-term deployments [7][8].

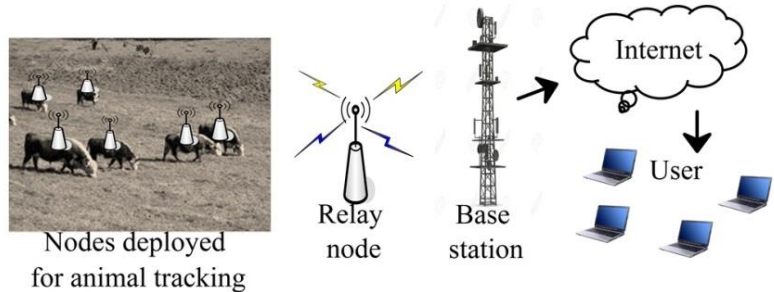
Figure 2: Animal tracking using WSN

In this paper, a method called a hybrid routing technique is recommended by comparing with conventional methods to improve network performance. The network is subdivided into various regions based on the distance of the sink from each sensor node. A relay node is placed within the network that can be recharged from time to time. Relay nodes are a special category of sensor nodes deployed in the network to share the load of overloaded nodes [9]. Since the relay node provides efficient data gathering to enhance the lifetime, the position of the relay should be engineered carefully. Figure 2 depicts the use of a relay node that is responsible for collecting data sensors deployed in livestock tracking [10] and transmitting the data to BS for further processing. The relay nodes balance the process of data gathering while maximizing the network lifetime and making the network fault tolerant. The relay node maintains, supports and recovers communication in the network [11]. The sensor nodes placed nearer to the BS or relay can transmit its data directly to them instead of forming clusters and CHs. The LEACH protocol can be used in specific regions only instead of the whole network to conserve energy in the unnecessary transmission of data.

The following sections of the paper can be elaborated as follows. Section 2 reviews the related published work on routing algorithms used in WSN. Section 3 presents the proposed network model description. Section 4 analyses the simulated results of the network performances and briefs a fair comparison of the proposed Hybrid algorithm with other existing protocols. And finally, section 5 indicates the future research followed by the conclusive remark.

## 2. Related Works

Researchers have been focusing on crucial issues in network designing such as enhancing network lifespan, effective power utilization and obtaining maximum throughput [12]. The position of the sink node and its mobility [13] has also been investigated as it plays an important role in communicating the data from the real world to the digital cloud through the Internet. A lot of advancement has been done with basic protocols and algorithms to achieve better network performance [14][15], either to lower power consumption and extend the life of sensor nodes or decrease the computational complexity to distribute load evenly in the network.

TL-LEACH [16] uses the concept of two CHs termed primary and secondary CHs. The protocol rotates the local cluster BS randomly. Similar to LEACH, CHs in tier 1 collects information from member nodes, on the other hand, the CHs in the second tier contains CHs of the first tier as a member node. This balances the network load, owing to which the hence network longevity increases. MR-LEACH [17], another energy efficient routing algorithm, conserves power by partitioning the field into numerous layers of clusters. Each layer CH communicates with adjacent layers for transmission of data to BS. The network field is slip into the uneven and unequal size of clusters in EEUC [18]. The cluster size varies with respect to BS; the ones nearer to BS are comparatively smaller to those that are far away. The energy efficient unequal clustering preserves energy during the inter-cluster communication of data. Network lifetime was also effectively enhanced in [19], where the candidate cluster head calculates its lifetime in a distributed manner. The energy-balanced clustering scheme selects CH efficiently with maximum residual energy for proper energy distribution.

M-GEAR [20], a gateway based routing protocol uses a special node placed within the network. The nodes placed close to gateway node send data directly to it instead of BS. The technique balances the load effectively in the network thereby increasing network lifetime. Multi-criteria based CH selection is discussed in [21] that enhances the network stability and lifetime by considering parameters like distance from neighboring nodes and center, power level, number of times a node has acted as CH, etc. the network is divided into several zones, each having its respective zone head selected efficiently to increase the network lifetime. A sub-netting based routing protocol (SNRP) [22] divides the network into regions based on the location of nodes. In case the BS is placed at a distant place away from the network, a rechargeable sub-BS is placed within the network that transfers data from nearby nodes and CHs. Simulation result shows the model yields better performance as compared to a network that follows only the LEACH algorithm.
For large scale sensor application, deployment of devices plays a major role in deciding the network performance. A hybrid deployment was discussed in [23], that balances connectivity and lifetime goals. In [24], authors have analyzed two deployment methods for relay node deployment in a heterogeneous WSN called lifetime-oriented and hybrid deployments. The analysis provides a trade-off between lifetime and connectivity in large scale networks and simulation shows better results.

The cluster head plays an important role in any clustering algorithm. Essential functions like fusion and processing of data are performed at the head level of the network which consumes extra energy as compared to sensing nodes [25]. The energy consumption is even larger for large scale scenarios where BS is located at a distant place. Replacement of the dead CHs with new ones adds to extra power consumption which may shorten the lifetime of the network. This article aims to design a hybrid routing algorithm that saves network energy by using a special node called relay within the network. The use of relay reduces the distance of transmission of data from CHs to BS. The sensor nodes are also devised with two different power levels with respect to its location in the network field. This distribution of nodes also helps in extending the network lifetime to a whole new level.

## 3. Proposed Work

A clustered hierarchically arranged network operates in three modes of data transmission [12]. The transfer of data from sensing nodes to respective CHs is the intra-cluster transmission. Communication between CHs is termed inter-cluster transmission. And finally, the CHs transmits the fused data to the BS or sink under long-haul transmission [26]. The power requirement in all these modes of operation varies significantly, for instance, intra-cluster communication requires a low level of energy as compared to other modes of transmission [26]. Also, the nodes near to BS will consume less energy in transmission as compared to ones located at a distant place. Hence a hybrid model is designed with sensor nodes having two different energy levels. The location of each type of node is fixed and cannot be altered.

Some reasonable assumptions considered to study the behavior of the proposed network are as follows:
   a. Sensor nodes are assumed to be quasi-static, i.e., they can move within the boundary of the specified region.
   b. There are two categories of sensors based on initial energy; normal and advanced nodes
   c. The position of BS and relay node is fixed. Both do not have any energy constraints.
   d. Nodes are deployed randomly within each region and transmit their data periodically in TDMA slots. The nodes turn on their radio only during the allocated time slot and switch it off thereafter to save power.
   e. The relay node receives the data from each CH and from Region 3 to forward it to BS.

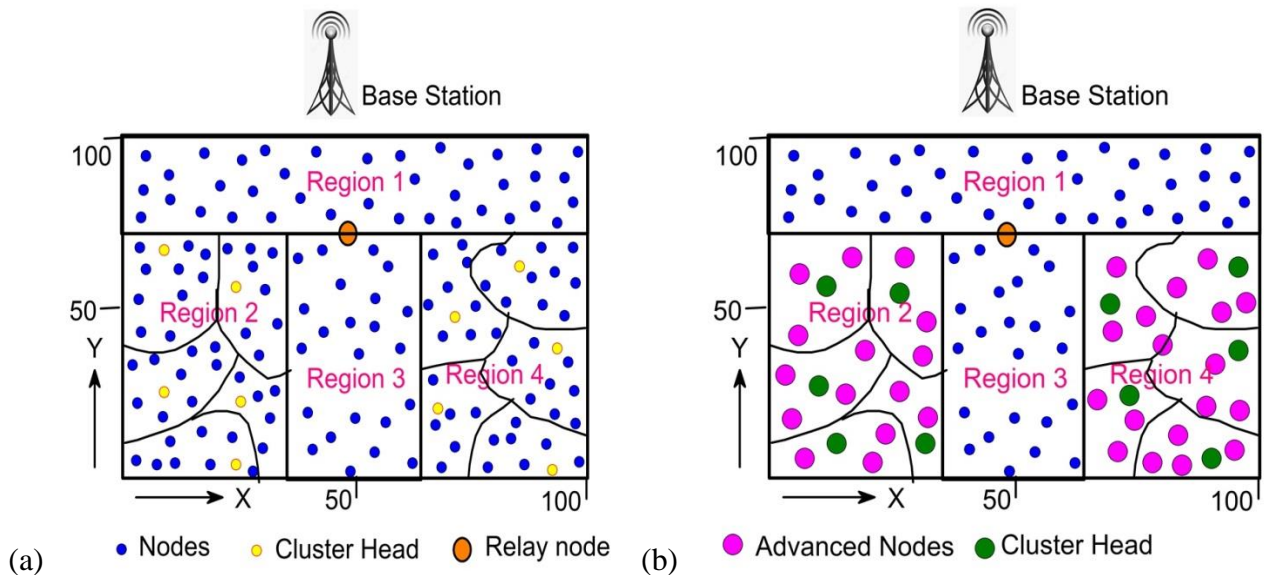

Figure 3. (a) Hybrid Homogenous Model (b) Hybrid Heterogeneous Model

Considering the network as farm automation, where animals need to be monitored on daily basis, the base station has to be located at a remote place [27]. Sensors need to be placed on all type of animals for monitoring their activities. The distance from BS for each section of livestock will vary according to its position. Hence, dividing the network into zones with respect to proximity will solve issues like load and power distribution. Figure 3(a) shows a hybrid structure introduced in [22] that divides the network into regions considering all nodes with equal energy. Figure 3(b) shows the proposed hybrid model with four small regions depending on the type of communication and the power level of the node. The structure also aims to reduce the average transmission distance by efficiently locating the relay node within the network. Clustering is done in Region 2 and 4 using standard LEACH protocol that comprises advanced nodes only. The normal nodes in Region 1 and 3 directly send its data to BS and relay respectively. The relay node allots TDMA slots to CHs in Region 2 and 4 to send their fused data that can be forwarded to BS. The relay collects all data from Region 2, 3 and 4 and performs data aggregation to remove redundancy and then transmits to BS. The data is then forwarded to the cloud and processed to make necessary required action [28].

The mode of communication in the proposed network is described as follows:
a. Node to BS as in Region 1: This direct communication consists of single hop as well as multi-hop communication.
b. Node to CH as in Region 2 and 4: Sensed data from nodes in these regions are sent to respective CHs in single hop transmission. The CHs collects data and then routes the fused data to the relay node for further communication.
c. Node to the relay node as in Region 3: Nodes in the region transmits the data directly to the relay node. The relay node shortens the transmission distance between the distant nodes and the sink by acting as a hop communicating them [9].
d. Relay node to BS: The relay receives all the data within the network and communicates it to the BS in a multi-hop manner by estimating a minimum routing path [29].

Consider '$n$' sensor nodes are randomly deployed in the network field with '$m$' percentage of advanced nodes. If $E_0$ is the initial energy of normal nodes then advanced nodes have $E_0(1+\alpha)$ energies. The symmetrical communication model shown in Figure 4 can be either a free space model or multi-path fading model [30] owing to the distance from sensing to the receiving node.

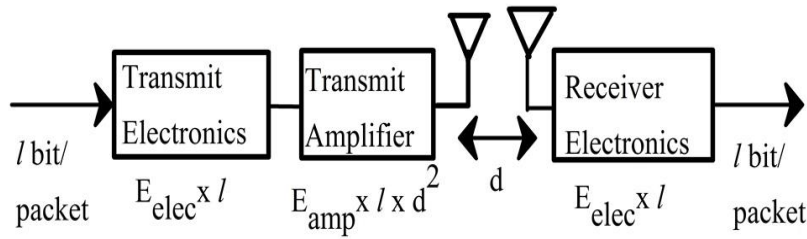

Figure 4. Radio Energy model

The energy consumption of a sensing node in transmitting '$l$' bits in a packet to another node some '$d$' meters apart can be given as [31]:

$$E_{Tx} = E_{Tx\_elec}(l) + E_{Tx\_amp}(l,d) \tag{1}$$

$$E_{Tx}(l,d) = \begin{cases} E_{elec} \times l + E_{fs} \times l \times d^2, d \leq d_0 \\ E_{elec} \times l + E_{amp} \times l \times d^4, d > d_0 \end{cases} \tag{2}$$

$E_{elec}$ is the energy consumption per bit by receiver or transmitter [26]. $E_{fs}$ and $E_{amp}$ are the parameters of amplification related to the free-space model and the multi-path fading model respectively. $d_0$ represents threshold distance given as the ratio between $E_{fs}$ and $E_{amp}$ as given by equation (3)

$$d_0 = \sqrt{\frac{E_{fs}}{E_{amp}}} \quad (3)$$

The energy consumed in receiving a '$l$' bits packet is given as:
$$E_{Rx}(l) = E_{elec} \times l \quad (4)$$

Energy consumed in direct transmission of '$l$' bits to relay located at distance $d_G$ from any node in Region 3 can be given as [22]:

$$E_{Tx\_G} = \begin{cases} E_{elec} \times l + E_{fs} \times l \times d_G^2, & d_G \leq d_0 \\ E_{elec} \times l + E_{amp} \times l \times d_G^4, & d_G > d_0 \end{cases} \quad (5)$$

where $d_G$ is the Euclidean separation of a node in Region 3 and the relay node.

In the initial round, all the nodes in Region 2 and 4 are eligible in the CH selection process. However, in subsequent rounds, the CH is selected by calculating the residual energy of each node and with the probability '$P_{adv}$'. A node can elect itself as a CH after every $r = 1/P_{adv}$ round.

The non-CH nodes in the current round are grouped in set '$S$'. At the beginning of each round, any node belonging to '$S$' randomly chooses a number in the range [0, 1]. Only if the number is found less than $T(n)$, the predefined threshold, then that node is declared to be the CH for that particular round

$$T(n) = \begin{cases} \dfrac{P_{adv}}{1 - P_{adv}(r \bmod \dfrac{1}{P_{adv}})}; & for\ all\ n \in S \\ 0; & Otherwise \end{cases} \quad (6)$$

After the CH of each region is elected, a control packet is transmitted using CSMA protocol [32] to the nodes. The nodes in return send ACK message and join to the nearest CH. The CH then assigns TDMA slots for the member nodes to transmit their sensed data. The node transfers data only during the allotted time period and shifts to sleep mode by turning off the radio in idle condition. Hence, the energy dissipated by the nodes decreases substantially.

## 4. Simulation and Result Analysis

100 sensor nodes are deployed over an area of 100×100 m² as shown in Figure 6. The value of 'm' is kept constant at 0.5 and α is selected as 1. The BS (red color) is located at coordinates (50,120) which is outside the sensing field. The position of relay nodes in the networks and optimized division of regions has a strong effect on network performance. Hence a brief analysis has been done to select the best position for the relay and divide the network accordingly into regions.

### A. *Relay Position*:

Since the BS is positioned in a distant place, the relay should be placed within the network in such a way that it should be able to communicate with all the nodes. Hence the optimum position should be at X= 50m. However, to select the Y coordinate, the network lifetime is analyzed considering four different cases such as Case 1 for (50, 00), Case 2 for (50, 20), Case 3 for (50, 50) and Case 4 for (50, 80). From Figure 5 we can deduce that Case 4 yields the maximum lifetime in both scenarios, i.e. 2400 and 4990 rounds for the homogeneous and heterogeneous models respectively. Hence the relay (green color) is placed within the network at coordinates (X, Y) = (50, 80).

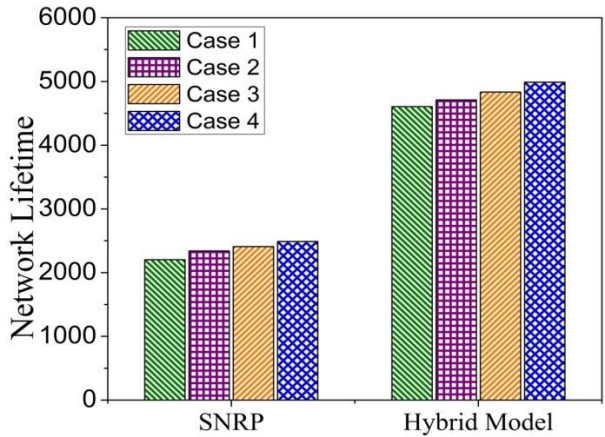
Figure 5: Relay Position

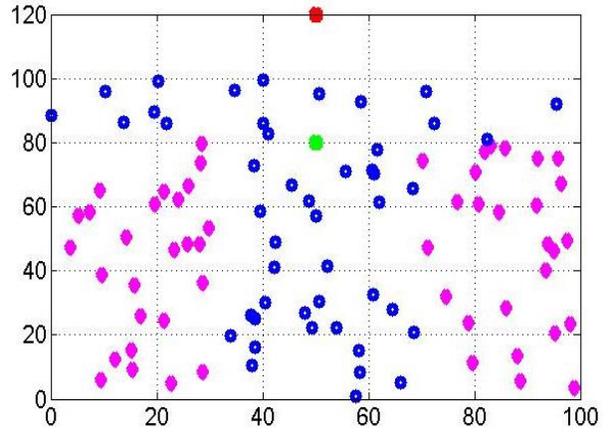
Figure 6. Deployment of Sensor Nodes

**B.** *Division of regions*:

After the placement of the relay node, dividing the network into small regions based on distance to BS or relay becomes a tedious task. In the proposed model, the nodes nearer to BS need to transfer the data directly, hence Region 1 is conventionally selected from 80 to 100 along Y-axis. Network Stability and lifetime are two major parameters in deciding the efficiency of any network. Figure 6 shows the analysis of possible division for Region 3; with five cases along X-axis, i.e. 10m to 90m, 20m to 80m, 30m to 70m, 40m to 60m and 50m to 50m. Figure 7(a) shows the number of rounds elapsed before the death of the first node in the network which is highest for the case 30m to 70m. Similarly, from Figure 7(b), it can be found that the last node dies out at 2429 and 4830 rounds for the third case. Hence Region 3 is defined between 30 to 70m along X-axis and 0 to 80m along Y-axis. For Region 2 and 4, the X-axis coordinates are 0 to 30 and 70 to 100m respectively and up to 80m for Y-axis.

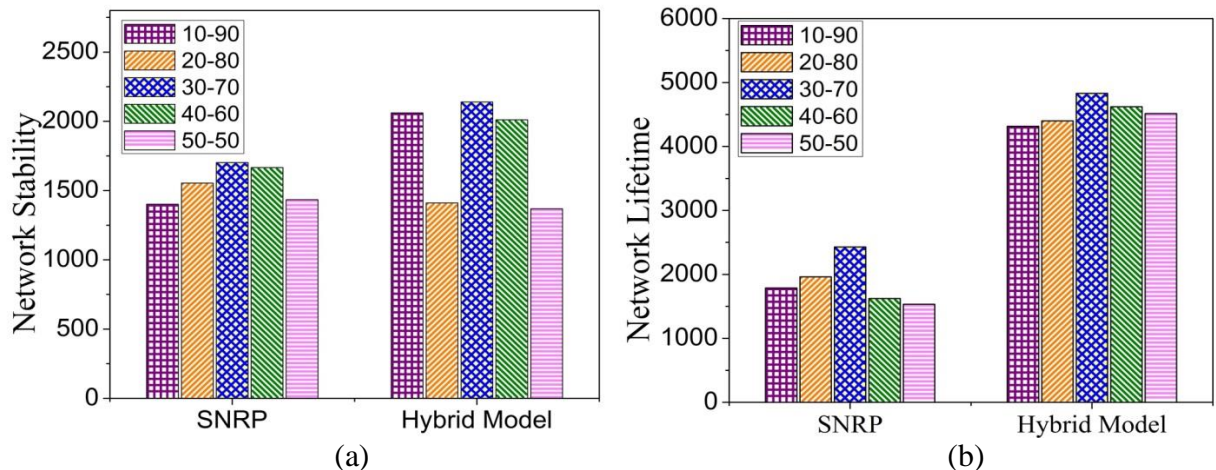
Figure 7. Region-wise Comparison (a) Network Stability (b) Network Lifetime

The system parameters used to simulate different metrics using MATLAB is given below in Table 1:

Table 1. System Parameters

| Symbol | Description | Value |
|---|---|---|
| $E_0$ | The initial energy of normal nodes | 0.5J |
| $E_{amp}$ | Energy dissipation: receiving | 0.0013/pJ/bit/m$^4$ |
| $E_{fs}$ | Energy dissipation: free space model | 10/pJ/bit/m$^2$ |
| $E_{da}$ | Energy dissipation: aggregation | 5pJ/bit |
| $E_{elec}$ | Energy consumed per bit | 5nJ/bit |
| $P$ | The probability of CH selection | 0.1 |
| $l$ | Packet size | 4000 bits |

The hybrid model with heterogeneous nodes is compared with LEACH, SNRP and SEP protocols to study the behavior of the network for metrics such as network lifetime, residual energy and throughput.

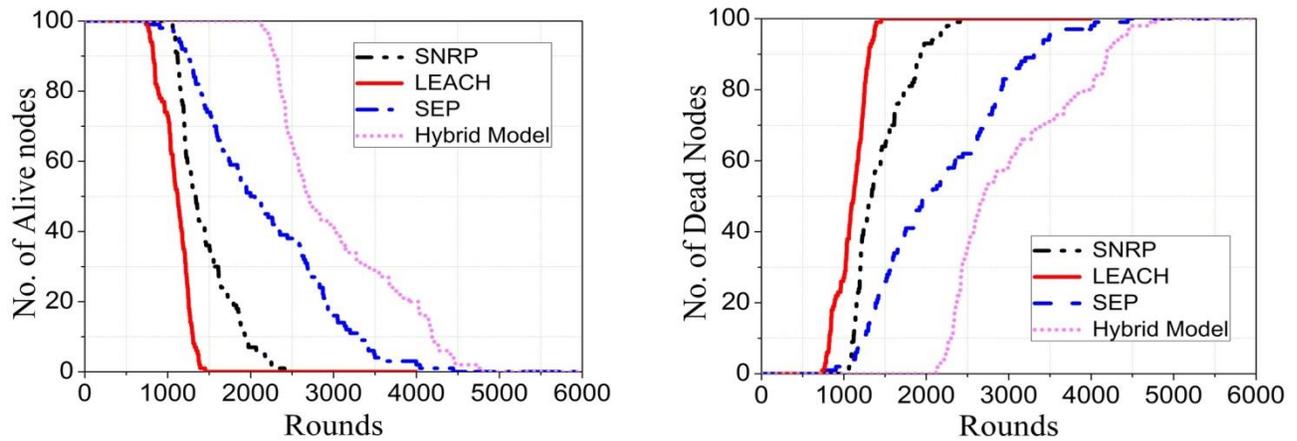

(b)

Figure 8. (a) Percentage of Alive nodes   (b) Percentage of Dead nodes

Figure 8 shows the performance comparison for two important parameters called lifetime and stability of the network. Stability period is calculated by the number of rounds the first node of the network takes to die and Lifetime is the total number of rounds of network operation. These two parameters help to decide the efficiency of any routing algorithm. Figure 8(a) clearly shows the Hybrid model has the maximum stability period which is 2118 rounds as compared to other protocols. For LEACH and SNRP which performs in a homogenous environment, the percentage of alive nodes is much less, i.e. 800 and 1000 rounds. SEP, on the other hand, is a heterogeneous routing algorithm; however, the election of CH is not restricted to any one type of node which leads to uneven energy distribution in the network. In the proposed hybrid model, clustering and CH has been restricted to only those regions that have a maximum distance from the BS. Similarly, the network lifetime is also enhanced for the proposed model which is up to 4778 rounds. This is clearly because of well-distribution of energy among sensor nodes. In LEACH and SEP, the nodes located at a longer distance consume more energy than those placed nearer to the BS. As a result, these nodes and CH deplete energy at a faster rate resulting in a breakage of the network. The hybrid protocol balances the energy consumption by placing a relay within the network for direct transmission of data. Also classifying the nodes according to supplied energy contributed to maintaining the energy balance in the network.

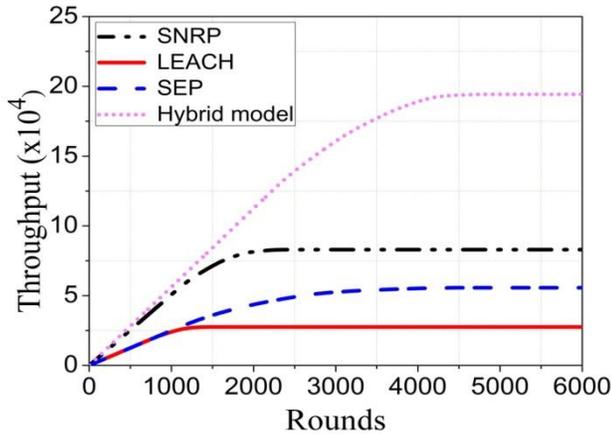

Figure 9. Throughput

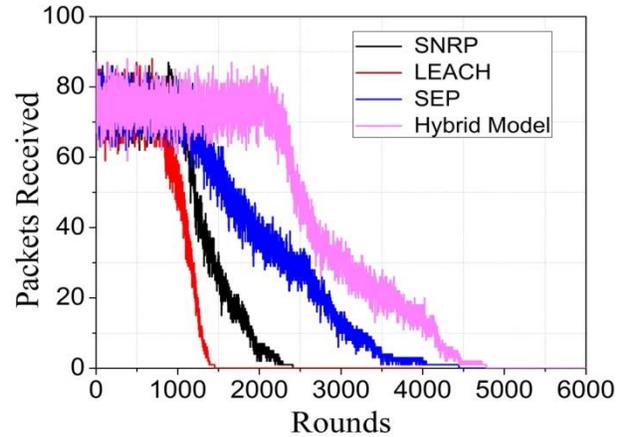

Figure 10. Data Packets Received

The ultimate goal of any network is to deliver maximum data to the base station with less packet drop ratio. Achieving a large data collection rate is a challenging task for WSN where the energy and communication resources are limited [33]. Figure 9 shows the rise in throughput of the network for a heterogeneous hybrid model in comparison to LEACH, SEP and SNRP routing protocol. The main reason behind this is the division of the network into regions owing to distance and energy. Advanced nodes in the specified regions stay alive for a longer time and are able to send more of the sensed data in the form of packets. The packet received ratio is shown in Figure 10. For LEACH, the packets from 70 to 85 nodes have been received for the first 900 rounds wherein the hybrid model, packets from the same number of nodes continues to be received till 1100 rounds. The value is even more in a heterogeneous environment where maximum data transmission occurs till 2500 rounds and then decreases gradually.

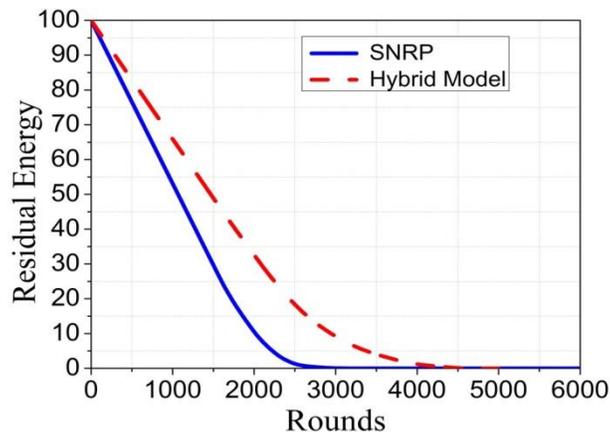

Figure 11. Residual Energy

Energy always remains the most constrained parameter in a wireless network. Figure 11 shows the comparison of average residual energy for homogeneous and heterogeneous models. The overall energy of the network is 100J, where the normal nodes are supplied with an initial energy of 0.5J. The value of α is made constant at 2, which indicates advanced nodes have 1.5J energy. In comparison to a network with nodes of the same energy level, the hybrid network performs well as far as residual energy is concerned. The energy level of each node is quite high when sensor nodes are assigned two different energy levels. The energy expenditure due to computation is also reduced since clustering is implemented in selected regions.

## 5. Conclusions

A single routing algorithm such as LEACH will not be able to deliver optimum network performance for scenarios where nodes once deployed could not be recharged at regular intervals. SNRP scheme reduced the network load by dividing into regions with a future perspective to implement a similar structure for a heterogeneous network. The proposed hybrid routing scheme was discussed for applications such as animal tracking, which extends network lifetime by choosing the mode of communication depending on the distance between sending and receiving node. Also, the sensor nodes are equipped with two different energy levels for better utilization of power to extend the network lifetime. The proposed method performs well in terms of throughput, network lifetime and energy consumption as compared to the network that follows clustering algorithms like LEACH and SEP as a whole.